\documentclass[10pt]{article}

\usepackage[framemethod=TikZ]{mdframed}
\usepackage{amsmath}
\usepackage{eqnarray,amsmath}
\usepackage{esvect}

\title{A comment on :'Tripartite entanglement versus tripartite nonlocality in 3-qubit GHZ-class states'}			
\author{Vibhu Gupta\footnote{vibhugupta01@gmail.com}}		
\date{\today}					

\begin{document}
\maketitle						

\section{Comment}

The paper [1] titled "Tripartite entanglement versus tripartite nonlocality in 3-qubit GHZ-class states", authored by Dr. Shohini Ghose and co-authors, published in PRL (Physical Review Letters 102, 250404 (2009)), does very important work on studying the nature of Svetlichny Inequalities (which are Bell-type inequalities) and sheds light on some important properties. It derives a result for the maximum expectation value of Svetlichny operator for GGHZ and MS states. In this comment, it is put forth that there is probably a mistake in the central result derived in the paper. \\

In the paper [1], a relationship is derived between maximum expectation value of a Bell type operator $S$ which is bounded by the inequality $|<S>| \leq 4$ [2]. A relationship is derived therein between maximum expectation value of $S$ and the 3-tangle  $\tau$, which quantifies tripartite entanglement. For the GGHZ state, the relationship derived in [1] is,

\begin{equation}
S(\psi_g) \leq \begin{cases} 4 \sqrt{1-\tau(\psi_g)}, \quad \tau(\psi_g) \leq 1/3\\
                                              4\sqrt{2\tau(\psi_g)}, \quad  \tau(\psi_g) \geq 1/3
                       \end{cases}
\end{equation}

For the GGHZ state, defined as $cos \theta_1 |000\rangle + sin \theta _1 |111\rangle$, the three tangle is $sin^2\theta_1$ [1]. The relationship for maximum operator value of Svetlichny operator, defined as $S= A(BK + B'K') + A'(BK'-B'K$), where $A= \vec{a}.\vec{\sigma_1}$,  $A'= \vec{a'}.\vec{\sigma_1}$, (similarly for B and C) and $K=C+C'$ and $K'=C-C'$ , and 3-tangle $\tau$ is derived in [1]. Doing a maximization of the Svetlichny operator  (defined as above), the paper [1] reaches to the equation (equation number 12 in [1]) which reads as:

\begin{equation}
\resizebox{0.9\hsize}{!}{$S(\psi_{g}) \leq \begin{cases} 4 cos 2\theta _{1} (cos^2 \theta _d + cos^2 \theta _{d'})^{\frac{1}{2}}, \quad  cos^2 2 \theta_1 (cos^2 \theta _d + cos^2 \theta _{d'} \geq sin^2 2 \theta_1 (sin^2 \theta_d +sin^2 \theta_{d'}) \\ 4 sin 2\theta _{1} (sin^2 \theta _d + sin^2 \theta _{d'})^{\frac{1}{2}} , \quad  cos^2 2 \theta_1 (cos^2 \theta _d + cos^2 \theta _{d'} \leq sin^2 2 \theta_1 (sin^2 \theta_d +sin^2 \theta_{d'})  \end{cases}      $}  
\end{equation}

which on simplification is shown in the paper [1] to reduce to,

\begin{equation}
S(\psi_g) \leq \begin{cases} 4 \sqrt{1-\tau(\psi_g)}, \quad \tau(\psi_g) \leq 1/3 \\
                                              4\sqrt{2\tau(\psi_g)}, \quad  \tau(\psi_g) \geq 1/3.
                       \end{cases}
\end{equation}
This equation (3) (equation number 14 in the paper [1]) is  wrong, as is shown by the argument which follows.  \\
First part of equation (2) implies that $S(\psi_{g}) \leq 4\sqrt{1-\tau(\psi _g)} $ when $ cos^2 2 \theta_1 (cos^2 \theta _d + cos^2 \theta _{d'} \geq sin^2 2 \theta_1 (sin^2 \theta_d +sin^2 \theta_{d'})$. Now simplifying $this$ condition we have 

\begin{eqnarray*}
cos^2 2 \theta_1 (cos^2 \theta _d + cos^2 \theta _{d'}) &\geq& sin^2 2 \theta_1 (sin^2 \theta_d +sin^2 \theta_{d'}),\\ 
\implies  cos^2 2 \theta_1 (cos^2 \theta _d + cos^2 \theta _{d'}) &\geq& sin^2 2 \theta_1 (2 - cos^2 \theta_d - cos^2 \theta_{d'}) \\
\implies cos^2 \theta _d + cos^2 \theta _{d'} &\geq& 2 sin ^2 2 \theta_1 \\
\implies 2 sin ^2 2 \theta_1 & \leq & 1 \\
\implies sin ^2 2 \theta_1 & \leq & 1/2\\
\implies \tau(\psi_g) &\leq& 1/2.
\end{eqnarray*}

(Because $cos^2 \theta _d + cos^2 \theta _{d'} \leq  1$.)  Similarly for the other inequality. Therefore the inequality (2) should actually be,

\begin{equation}
S(\psi_g) \leq \begin{cases} 4 \sqrt{1-\tau(\psi_g)}, \quad \tau(\psi_g) \leq 1/2 \\
                                              4\sqrt{2\tau(\psi_g)}, \quad  \tau(\psi_g) \geq 1/2.
                       \end{cases}
\end{equation}

I think this is a significant correction to the paper, as some experimental works have tried to fit their data according to the derived equation (3) in [1], which should actually be corrected to equation (4).

\section{References} \nonumber
1. Ghose et. al.  Physical Review Letters 102, 250404 (2009). \\
2. G. Svetlichny, phys. Rev. D. 35, 3066 (1987).

\end{document}